\documentclass[12pt]{article}
\usepackage{graphicx}
\usepackage{caption}
\usepackage{subcaption}
\usepackage{amsfonts}
\usepackage{amsmath}
\usepackage{amssymb}
\usepackage[margin=2.54cm]{geometry}

\begin{document}

\title{Geometric theory of inversion and seismic imaging}
\author{August Lau and Chuan Yin}

\maketitle

\begin{abstract}
The goal of inversion is to estimate the model which generates the data of observations with a specific modeling equation.

One general approach to inversion is to use optimization methods which are algebraic in nature to define an objective function.   This is the case for objective functions like minimizing RMS of amplitude, residual traveltime error in tomography, cross correlation and sometimes mixing different norms (e.g. L1 of model + L2 of RMS error).

Algebraic objective function assumes that the optimal solution will come up with the correct geometry.   It is sometimes difficult to understand how one number (error of the fit) could miraculously come up with the detail geometry of the earth model.    If one models the earth as binary rock parameters (only two values for velocity variation), one could see that the geometry of the rugose boundaries of the geobodies might not be solvable by inversion using algebraic objective function.

\end{abstract}

\section*{Introduction}

Geometric objective function is an alternative criterion for inversion.  The geometric theory of inversion attempts to use measures which are geometrical or topological in nature.   Topology is “flexible” geometry and the two are closely related.  Topology is like generalized geometry.  So when geometry is used in the paper, it could mean geometry or topology.   

This paper uses geometric optimization without prior information.  Just for clarification,  we are not constraining the structure and we are not constraining the gradients by using geologic dips or geologic interpretation.  

In our previous papers, we used Betti numbers as an objective function to measure the geometric simplicity.  Betti numbers could illustrate the geometric theory of inversion.

\section*{Practical Applications  : End game}

Instead of going through theory and then applications,   we will start with the end game in mind.  In seismic imaging,   the end-users are interpreters.  They take the seismic images and other information to generate prospects to drill for oil and gas.  

The end game is to find oil and gas.  Interpreters care a lot more about the geometric shapes which determine how they map the geologic structures.  The geometric shape determines the container which is the rock mass for oil and gas exploration.  Faults are also important geometric shapes which determine the geologic boundary between two geologic fault blocks.

\section*{Inversion Theory}

\textit{"With four parameters I can fit an elephant, and with five I can make him wiggle his trunk”} - attributed to John von Neumann 
\\

This is a humorous quote which has insight into the demise of inversion theory.

Let us consider Parseval's theorem. If $d$ is a one-dimensional function in time like a one-dimensional time series,  then Parseval's theorem states that 
 $$RMS (d(t)) = RMS (D(f))$$ 
up to a multiplicative constant, where RMS is root-mean-square, $d$ is the time data trace and $D$ is the frequency transform trace, $t$ is in time and $f$ is in frequency.

Any inversion algorithm error in time will mean that the inversion will have error in frequency domain for every frequency.   Certainly, some frequency error will be larger than other frequency error.   

The geometric interpretation of Parseval's theorem would mean that sinusoids will appear in the seismic data in time/depth since the error is the same in frequency domain.  It would also cause cycle skipping which means that the time data could line up in a different cycle for that frequency error.

The algebraic objective function could not escape the non-geologic appearance of sinusoids in the inversion result.   The algebraic objective function could also cause cycle skipping since each frequency could fit the data just as well and yet it could line up with the wrong cycle in time.

\section*{Synthetic Data Examples}

Let us turn to geometric theory of inversion.   We will start with a quick review of Betti numbers which describe the geometric/topological simplicity.  Betti numbers are computationally demanding.  

Figure 1 shows various constant velocity migrated images of 3 diffractions.   When the correct migration of 1500 m/s is used,  the diffractions are collapsed into 3 distinct impulses.   When we calculate the Betti numbers B1,  the Betti numbers drop to a low value when 1500 m/s is reached.   For real data,,  the Betti numbers will fall in a low range of values rather than a distinct minimum like synthetic model.  

Figures 2 through 10 show two more synthetic data examples, with triangular velocity anomalies. The imaging results show that when a smoothed or non-exact velocity model is used for the migration, the resulting image yields higher apparent Betti number. These further suggest that geometric measure, such as the Betti numbers, can serve as an optimization criterion.

The Betti numbers give us a geometric/topological objective function.   In real data,  whether it is algebraic/geometric objective function,  it could only give a range of low values and could never be a true minimum.   We have attempted to include more probability into the mix  to make it more precise. But it will not be precise because the “noise” is geological and not numerical.   

The real solution is to INTERPRET the result geologically.   Also INTERPRET the complex part of the data (e.g. residuals,  see below).

\section*{Importance of Residuals}

When we have the “best” solution to the seismic problem,  we should forward model the synthetic data from the best solution and see how well it matches the recorded data.   The difference is the residual.  In general,  we do not study the residuals.   But it is important to INTERPRET the residuals.  

Residuals are not random noise but have geometric meaning and geologic meaning.  They should be interpreted just like interpreting seismic image and seismic model.   

Also the residual tells us what is unsolvable given the inversion method.   We have used the terminology of “complex part” of the data in previous papers. Residual is one way to define the complex part which is the mis-match of modeled data and recorded data.

\section*{Interpretation of seismic imaging and cycle skipping}

One of the most prevalent problem in tomography and inversion is cycle skipping.   Both methods generate a velocity model which could then be used to migrate the seismic data.   Migration is an imaging algorithm which comes up with a seismic volume or seismic section in 3-D or 2-D.

Cycle skipping is “obvious” to an experienced interpreter.  We could detect cycle skipping by just “looking” at the seismic image.  But cycle skipping is a geometric concept which does not have a definition at this time.   If cycle skipping could be defined,  then we might be able to come up with an  objective function to minimize cycle skipping.

How do we generate good inversion algorithms and objective functions when we could not define the geometric problem of cycle skipping ?   Betti numbers are our attempt to minimize cycle skipping.

\textit{"There's no sense in being precise when you don't even know what you're talking about"}   - attributed to John von Neumann

\section*{Conclusion}

There are other geometric objective functions which we did not include in the paper, e.g., cohomology theory, network topology, counting branches for complex stratigraphy, rigidity to make long wavelengths more like linear segments, etc.  

We also have not mentioned the geometry of operators.  Typical geometry of  operators is like convexity and its relationship to L1 norm and sparsity.  The operators could be defined by differential equations which connect the model to the data.  Differential equations are not rigid but could be changed.  This is another level of abstraction compared to model space or data space.  Geometry of operators will be a different research direction to tie it to interpretation.

It would be our hope that algebraic objective function and geometric objective function will be combined to form a more general inversion theory in the future.  We prefer not to make prior assumptions like geologic dips or gradients which already bias the inversion.   We also prefer not to arbitrarily restrict illumination which could bias the geologic dip direction.  These constraints could lead to interpretation pitfalls of seismic data.  

Instead of making more geologic assumptions, it might be better to consider merging the geometric and algebraic objective functions.  The combination will maximize the data driven inversion.  As a last resort, we might have to make prior geologic assumptions.

Zhou (2006) conducted research on "deformable layer tomography" which is a forerunner of geometric theory of inversion. Our paper added a topological criterion like Betti numbers. Dependent on seismic interpretation objective, one might need quantitative requirement like geometry or qualitative requirement like topology.

\bibliographystyle{plain}
\bibliography{geometric_inversion2015}   

\begin{figure}
\centering
  \includegraphics[width=5in]{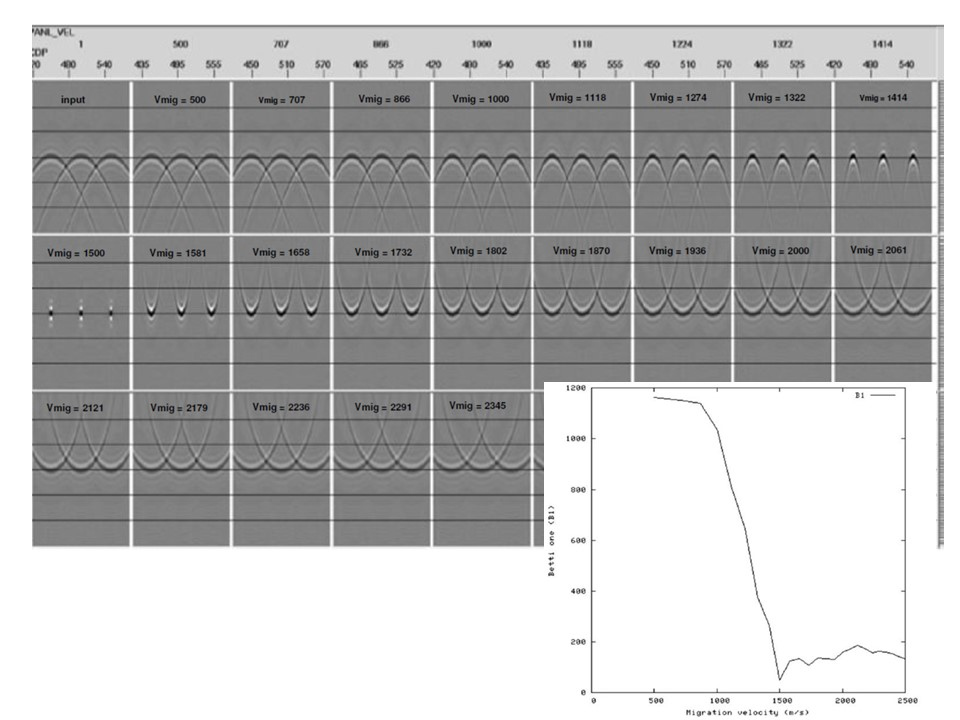}
\caption{Migrated images with various constant velocities and the corresponding Betti numbers.}
\label{fig1}
\end{figure}

\begin{figure}
    \centering
    \begin{subfigure}[b]{0.45\textwidth}
      \includegraphics[width=\textwidth]{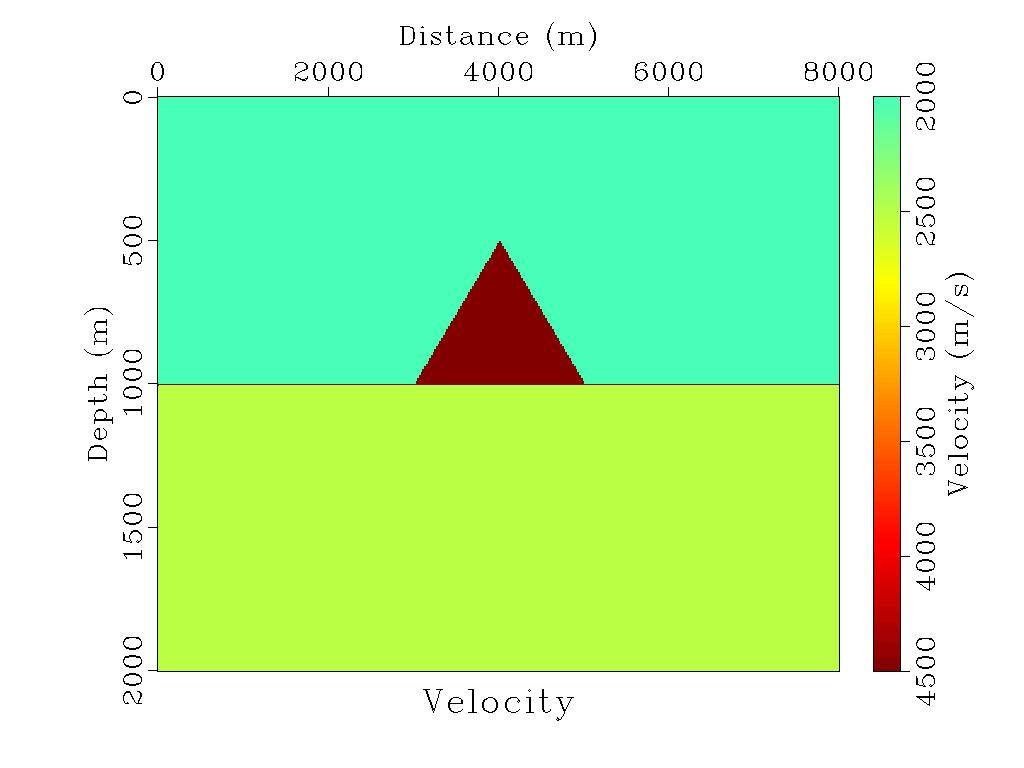}
    \caption{A toy velocity model}
    \label{vel}
    \end{subfigure}
    \hfill
    \begin{subfigure}[b]{0.45\textwidth}  
      \includegraphics[width=\textwidth]{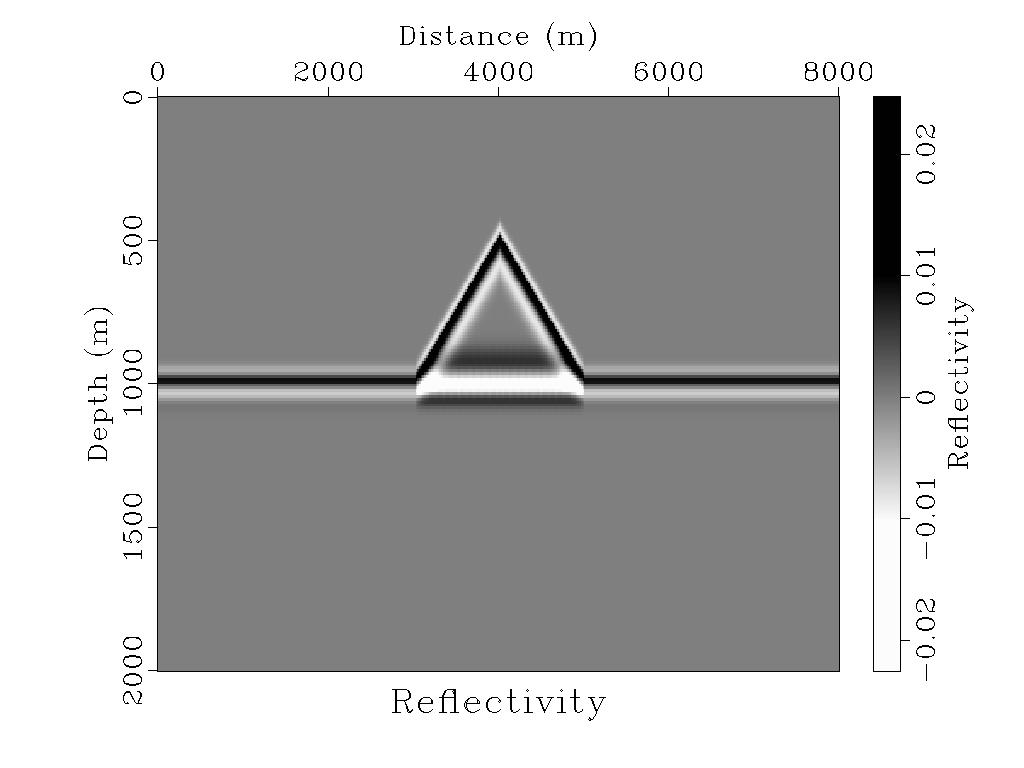}
    \caption{Reflectiviy of the velocity model}
    \label{ref}
    \end{subfigure}
    \caption{Velocity Model1 for imaging test.}
    \label{fig2}
\end{figure}

\begin{figure}
    \centering
    \begin{subfigure}[b]{0.45\textwidth}
      \includegraphics[width=\textwidth]{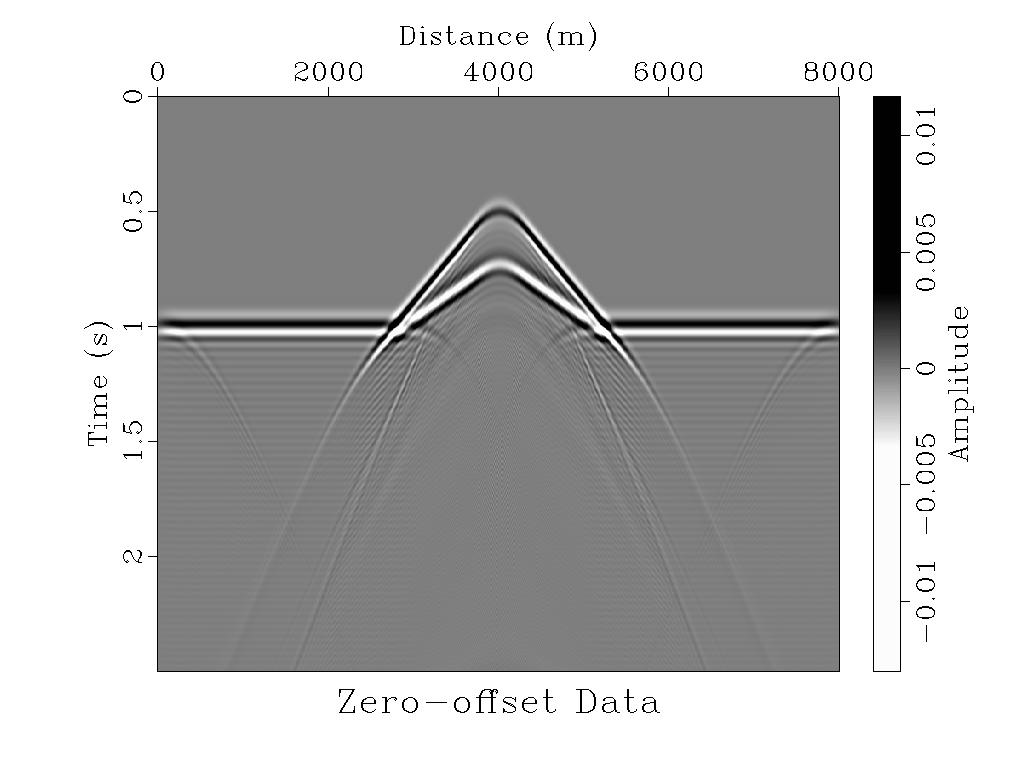}
    \caption{Modeled zero-offset data}
    \label{dat}
    \end{subfigure}
    \hfill
    \begin{subfigure}[b]{0.45\textwidth}
      \includegraphics[width=\textwidth]{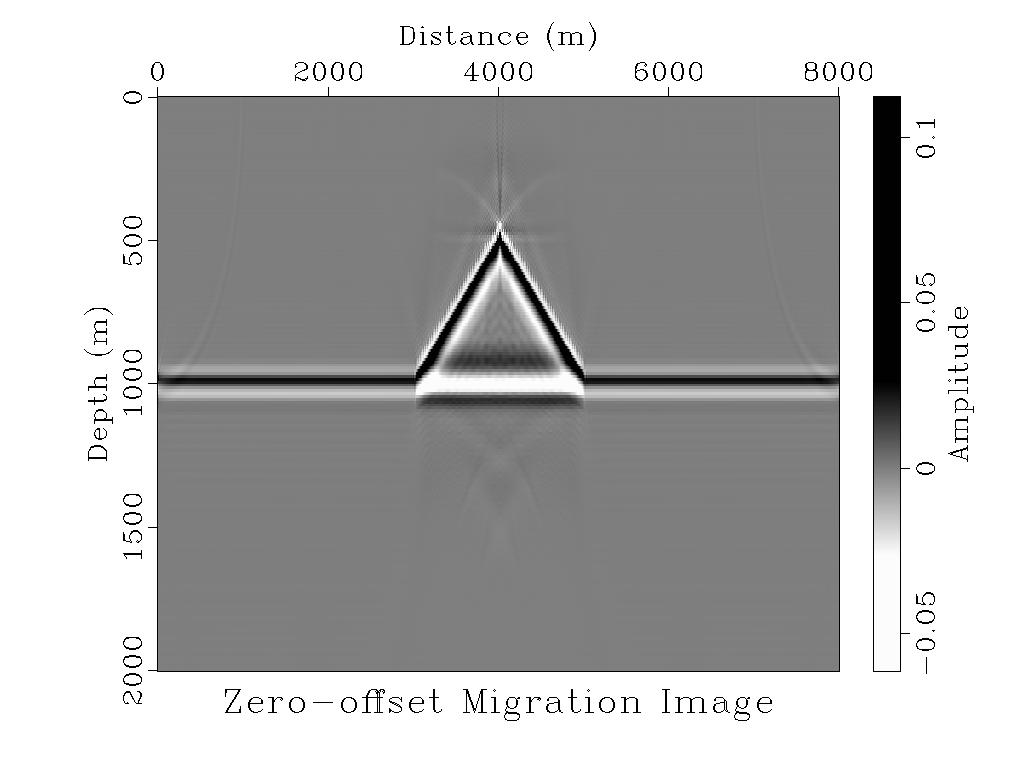}
    \caption{Migrated image using exact velocity}
    \label{img}
    \end{subfigure}
    \caption{Modeled zero-offset recording and migration result using exact velocity model (Betti number $B1 = 1$ with one "hole").}
    \label{fig3}
\end{figure}

\begin{figure}
\centering
  \includegraphics[width=4in]{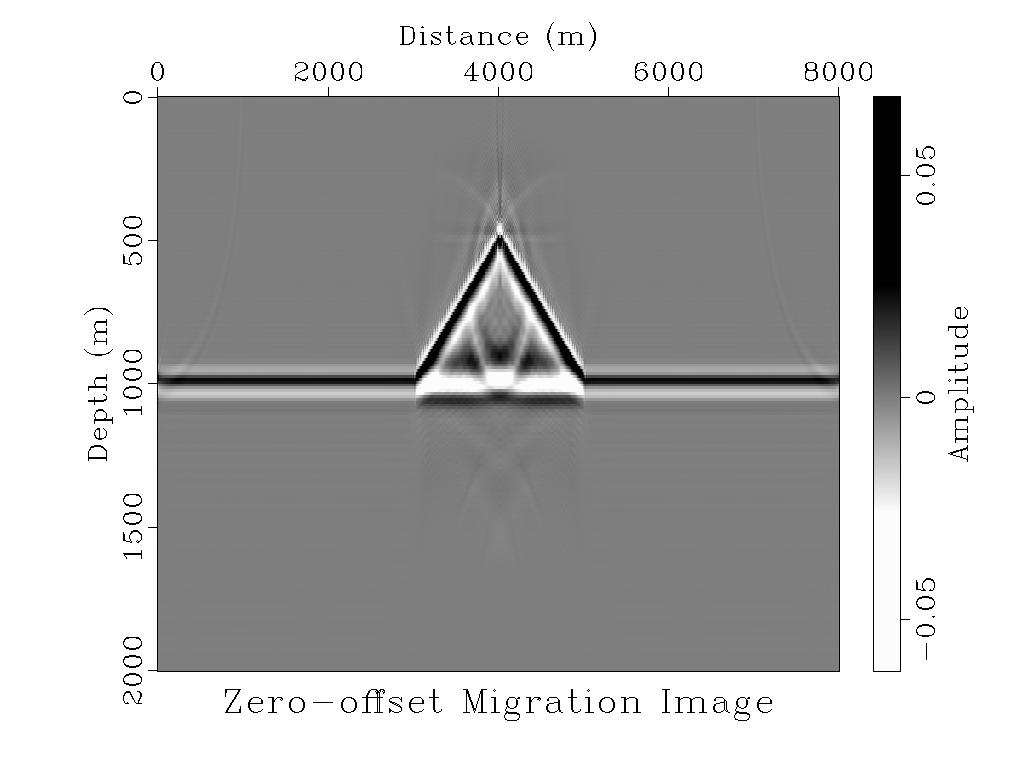}
\caption{Migrated image using a smoothed velocity model with a "rounded top" (Betti number $B1 = 3$ with three "holes")}
\label{imgm}
\end{figure}

\begin{figure}
\centering
  \includegraphics[width=6.5in]{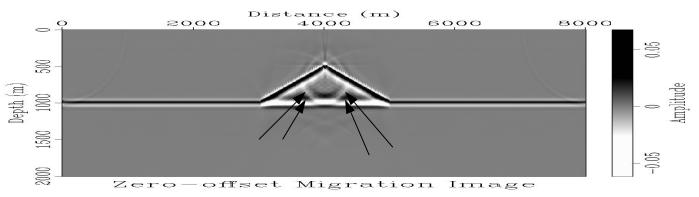}
\caption{Migrated image using a smoothed velocity model with a "rounded top" (Betti number $B1 = 3$ with three "holes"). With a close to 1:1 plot ratio, notice the "cycle skipping", i.e., one event split into two events within the triagular velocity anomaly zone.}
\label{imgm1to1}
\end{figure}

\begin{figure}
\centering
  \includegraphics[width=4in]{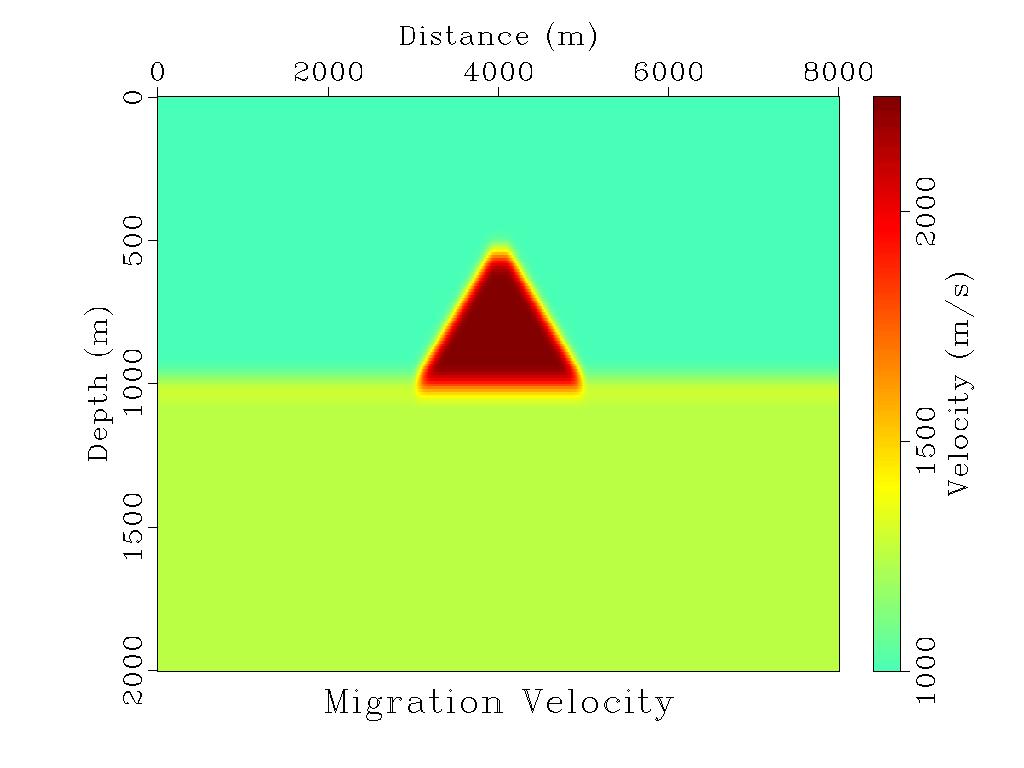}
\caption{A smoothed migration velocity field. This is to simulate a "realistic" solution with inversion and/or tomography using the $L2$ -norm.}
\label{migvelm}
\end{figure}

\begin{figure}
    \centering
    \begin{subfigure}[b]{0.45\textwidth}
      \includegraphics[width=\textwidth]{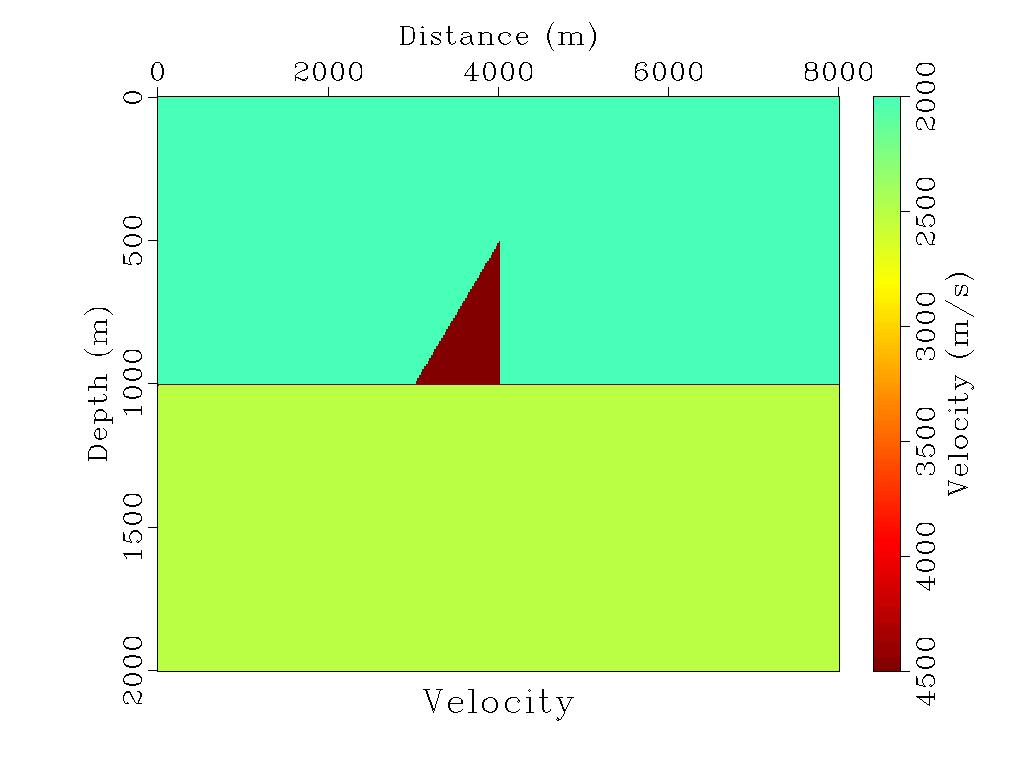}
    \caption{A toy velocity model}
    \label{m3vel}
    \end{subfigure}
    \hfill
    \begin{subfigure}[b]{0.45\textwidth}
      \includegraphics[width=\textwidth]{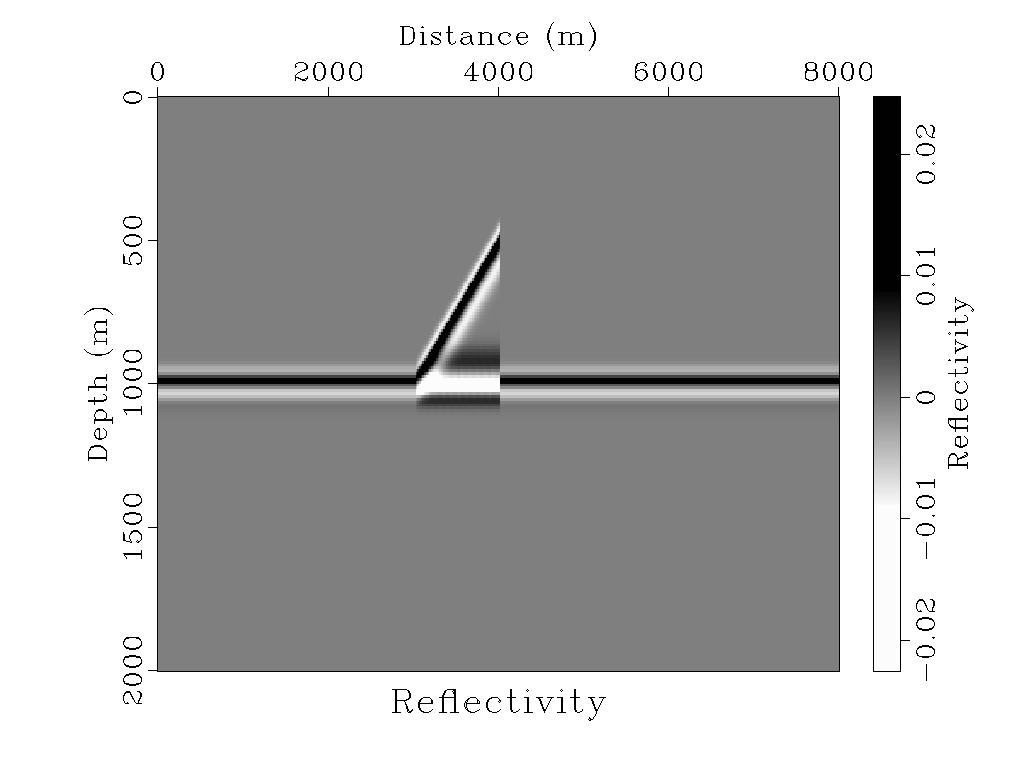}
    \caption{Reflectiviy of the velocity model}
    \label{m3ref}
    \end{subfigure}
    \caption{Velocity Model2 for imaging test.}
    \label{fig7}
\end{figure}

\begin{figure}
    \centering
    \begin{subfigure}[b]{0.45\textwidth}
      \includegraphics[width=\textwidth]{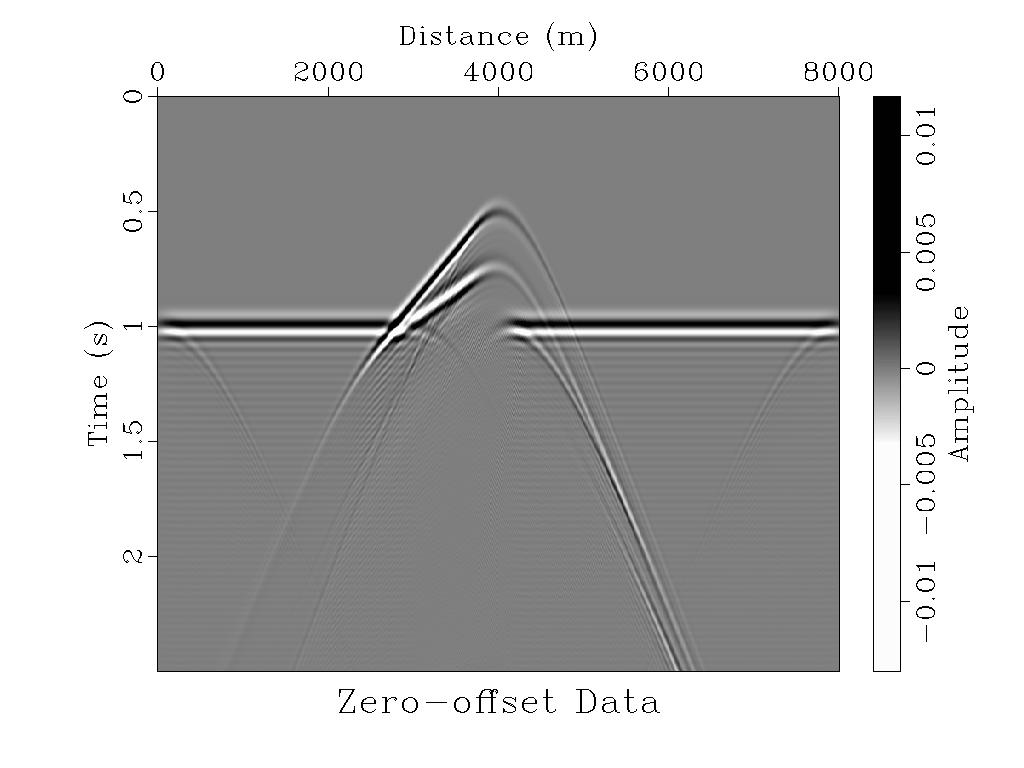}
    \caption{Modeled zero-offset data}
    \label{m3dat}
    \end{subfigure}
    \hfill
    \begin{subfigure}[b]{0.45\textwidth}
      \includegraphics[width=\textwidth]{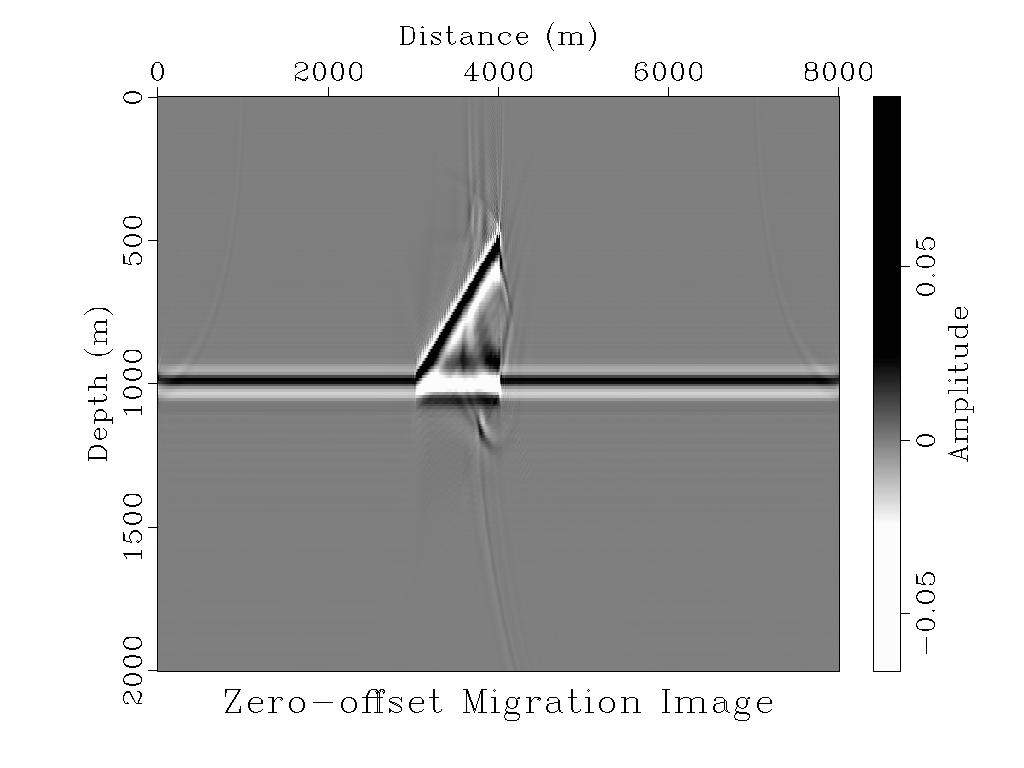}
    \caption{Migrated image using exact velocity}
    \label{m3img}
    \end{subfigure}
    \caption{Modeled zero-offset recording and migration result using exact velocity model (Betti number $B1 = 1$ with one "hole").}
    \label{fig8}
\end{figure}

\begin{figure}
\centering
  \includegraphics[width=4in]{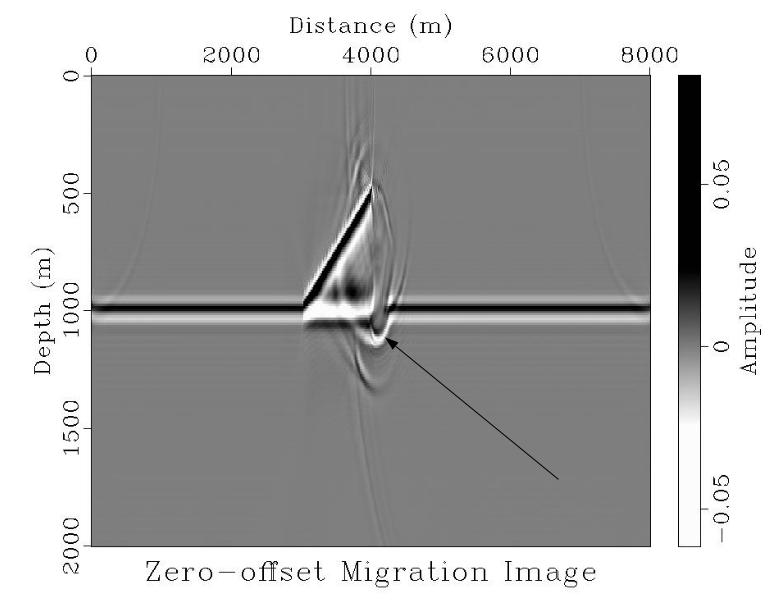}
\caption{Migrated image using a smoothed velocity model with a "rounded top" (arrow points to a curved cycle which is an artifact).}
\label{m3imgm}
\end{figure}

\begin{figure}
\centering
  \includegraphics[width=4in]{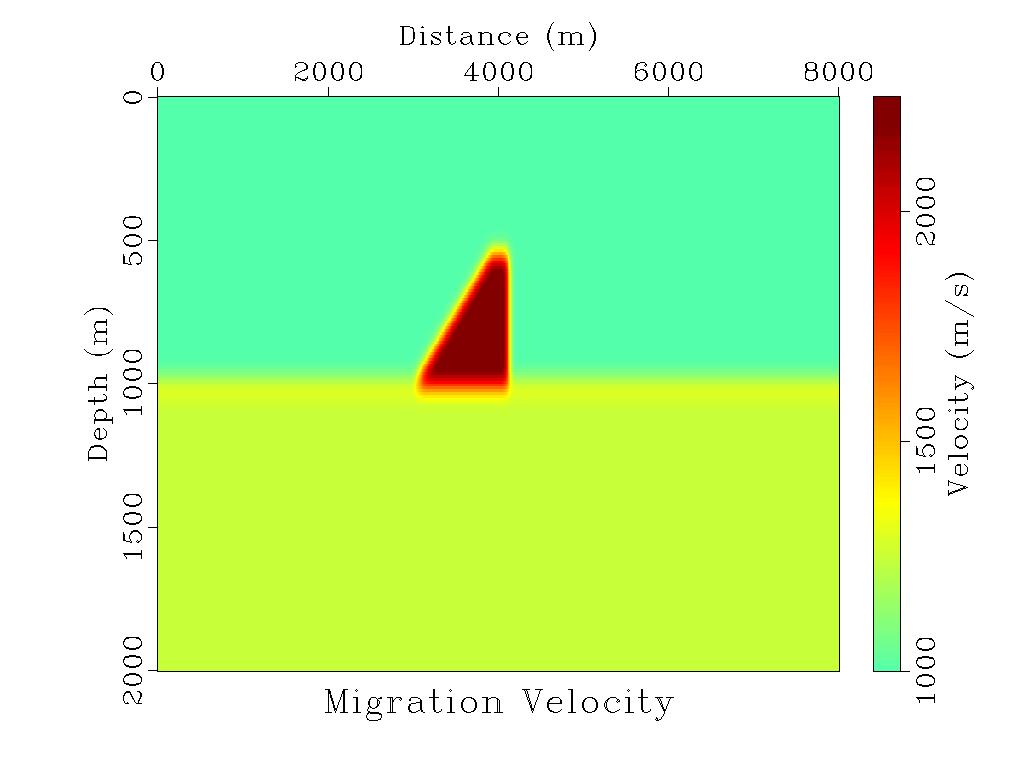}
\caption{A smoothed migration velocity field.}
\label{m3migvelm}
\end{figure}

\end{document}